# Pairing Effect on Spectral Statistics of Even and Odd Mass Nuclei

H. Sabri[a†], B. Rashidian Maleki[a], J. Fouladi[a] , N. Fouladi[a], M. A. Jafarizadeh[b,c]

[a] Department of Nuclear Physics, University of Tabriz, Tabriz 51664, Iran.

[b] Department of Theoretical Physics and Astrophysics, University of Tabriz, Tabriz 51664, Iran.

[c] Research Institute for Fundamental Sciences, Tabriz 51664, Iran.

† E-mail: h-sabri@tabrizu.ac.ir

## Abstract

The interplay of pairing is explored for the spectral statistics of nuclear systems with emphasis on the nearest neighbor spacing distributions by employing the kernel density and maximum likelihood estimation techniques. Different sequences prepared by all the available empirical data for low-lying energy levels of even and odd-mass nuclei in the $34 \leq A < 206$ mass region. A deviation to more regular dynamics is apparent for even-mass nuclei in compare to odd-mass ones, and there are suggestions of effects due to unclosed proton shells on more chaotic dynamics.

## Introduction

The investigations of spectral statistics and non-linear dynamics in different systems are receiving considerable attentions in the four past decades. Random matrix Theory (RMT) as the most commonly used tool in the investigation of the fluctuation properties of quantum system's spectra, describes a chaotic system by an ensemble of random matrices subject only to the symmetry restrictions [1-13]. Systems with time reversal symmetry such as atomic nuclei are described by Gaussian Orthogonal Ensemble (GOE). On the other hand, systems which classically integrable, i.e. non-chaotic, have been found to well characterized by Poisson distribution [3-6]. More recent developments in the statistical investigations have focused on the shell effects due to the dissipation and the breaking of dynamical symmetries while Guhr *et al* have considered the effect of breaking the isospin symmetry on spectral statistics [14].

The Nuclear Shell Model is the best available theoretical tool for calculating the properties of the low-lying states. Systematic studies for the spectral statistics using the Shell Model have been made in the $sd$ shell [15] and in the $pf$ shell [16]. The results in the $sd$ shell showed a general agreement with the Random Matrix Theory [15]. However, exceptions were found in the spectral statistics of nuclei in the $pf$ shell region [16-18] and in the $Pb$ region [19-20]. The dynamics in the low-energy region of semi-magic nuclei in these zones of the nuclear chart were close to regularity. The difference in the intensity of the $T=1$ and the $T=0$ residual interaction was considered to be the main responsible for that behavior [16]. The $T=1$ residual interaction, the only part that is active on semi-magic nuclei, is not strong enough to perturb completely the regular motion in the mean field the increment or decrement of the level repulsion was very dependent on the intensity of the mean field.

The pairing effect on the spectral statistics of nuclear systems have been described theoretically by Molina *et al* in Ref.[21-22] where by changing the intensity of the pairing interaction with respect to the realistic residual interaction, they have suggested, the pairing interaction controls the spectral statistics of low-lying levels. Also, their results proposed that, an increase in the intensity of the pairing interaction drives the statistical properties of the spectra of low-lying states closer to Poisson. This mean, one can expect a more regular dynamic for even-mass nuclei in compare to odd-mass ones in this region of energy due to the mean field effects [21].

In this work, we consider the spectral statistics of different sequences which prepared by even- and odd-mass nuclei. With using all the available experimental data [23-25], i.e. $2^+$ and $4^+$ levels of even- and also $1/2^+$, $3/2^+$, $5/2^+$ levels of odd-mass nuclei in which the spin-parity $J^\pi$ assignment of at least five consecutive levels are definite, levels are combined in several ways to search for effects due to mass, the intensity of pairing and also the types of pairs on the spectral statistics. Also, we have used the Maximum Likelihood (ML) [26] and Kernel Density (KD) estimation [27-28] techniques to consider the spectral statistics of sequences with high accuracy by both parametric and non-parametric estimation methods, respectively.

This paper is organized as follows: Section 2 briefly summarizes the theoretical aspects of pairing Hamiltonian and data sets which have used in this analysis. Section 3 dealt with reviewing statistical approaches containing unfolding processes, MLE and KDE techniques and finally, section 4 contains the numerical results. Section 5 is devoted to summarize and some conclusion based on results given in section 4.

## 2. Data set

In this section, we describe our choice of levels in even and odd-mass nuclei. In nuclear physics, pairing and quadrupole-quadrupole interactions can be regarded as the most important interactions. This concept was proposed by Racah as a seniority scheme in atomic physics [29-31]. Pairing is regarded as a simple and most regular part of nuclear interaction. In the low-lying part of nuclear spectra, it yields a pair condensate that influences strongly on all nuclear properties. On the other hand, according to the standard BCS description borrowed from the macroscopic theory of superconductivity, the excitation of the system breaks pairs, removing them from the interaction domain and blocking the scattering phase space for remaining pairs. Then, at some excitation energy ~ or temperature a sharp second-order phase transition occurs to a normal-heated Fermi liquid where the pairing effects are usually neglected [30]. The thermodynamically properties, entropy and *etc* [32-33] have been studied by different authors. The statistical properties of low-lying energy levels can be used for investigating the pairing effect on nuclear structures, too.

To study the obvious pairing effect in Hamiltonian, some authors use the following Hamiltonian [21],

$$H = H_{real} + H_{Pair}(G) \quad , \tag{2.1}$$

Where $H_{real}$ is a realistic Hamiltonian and

$$H_{Pair} = -G\sum_{j,j'}\sum_{m,m'>0}(-1)^{j+m}(-1)^{j'+m'}a^{\dagger}_{jm}a^{\dagger}_{j,-m}a_{j',-m'}a_{j',m'} \quad , \tag{2.2}$$

In the theoretical descriptions, pairing was considered as a part of nuclear Hamiltonian and by changing the intensity of the pairing interaction, $G$, the statistical properties have been analyzed in some special nuclei [21-22]. These analyses suggested, the pairing interaction controls the spectral statistics of low-lying levels and also the variation of the strength, changes substantially the fluctuation properties of low-energy levels. It means, an increase in the intensity of the pairing interaction drives the statistical properties of the spectra of low-lying states closer to Poisson, i.e. more regular statistics. On the other hand, a reduction of the pairing interaction in the realistic force proposed a more similar to the GOE properties in the spectral statistics.

To consider similar investigation via experimental data, some sequences constructed by all the available empirical data taken from Refs.[23-25]. We have followed the same method given in Ref.[11], namely, we selected nuclei in which the spin-parity $J^{\pi}$ assignments of at least five consecutive levels are definite. In cases where the spin-parity assignments are uncertain and where the most probable value appeared in brackets, we admit this value. We terminate the sequence in each nucleus when we reach at a level with unassigned $J^{\pi}$. We focus on $2^+$ and $4^+$ levels of even and $1/2^+$, $3/2^+$, $5/2^+$ levels of odd-mass nuclei (for their relative abundance in the specified nuclei).

## 3. Method of analysis

The fluctuation properties of nuclear spectra have been considered by different statistics such as Nearest Neighbor Spacing Distribution (NNSD) [1-4], the Dyson-Mehta $\Delta_3(L)$ statistic [12-14] and *etc*. To perform a statistical analysis for the NNS distribution of spherical nuclei in different mass regions, similar to every statistical analysis which using RMT, we must have a sequences of unit mean level spacing. This requirement is done by fitting a theoretical expression to the number $N(E)$ of level below the excitation energy $E$ which is regard as unfolding procedure. The expression used here is the constant-temperature formula [11]

$$N(E) = N_0 + \exp(\frac{E - E_0}{T}) \quad , \quad (3.1)$$

The three parameters $N_0$, $E_0$ and $T$ obtained for each nucleus vary considerably with mass number. Nevertheless, all three show a clear tendency to decrease with increasing the mass number $A$, using 2nd order polynomial function for each the three unfolding parameters. With excluding the available empirical data for five nuclei from all considered nuclei (114 nuclei), i.e, $^{44}Ca$, $^{110}Cd$, $^{120}Sn$, $^{130}Te$, $^{199}Hg$ which are semimagic or at subshell closure, we achieved the following values for each fitting parameters,

$$T = (3.82 \pm 0.19) - (0.009 \pm 0.003)A + (4.38 \pm 0.21)\times 10^{-5} A^2 \quad , \quad (3.2a)$$

$$E_0 = (3.21 \pm 0.11) - (0.020 \pm 0.008)A + (2.08 \pm 0.33)\times 10^{-5} A^2 \quad , \quad (3.2b)$$

$$N_0 = (0.75 \pm 0.08) - (0.010 \pm 0.020)A + (2.52 \pm 0.41)\times 10^{-5} A^2 \quad , \quad (3.2c)$$

With these quantities for each nuclei, we obtained the best fit for $N(E)$ which denote by $F(E)$. Now, the corrected set of energies is generated by means of [11]

$$E_i^{'} = E_{\min} + \frac{F(E_i) - F(E_{\min})}{F(E_{\max}) - F(E_{\min})}(E_{\max}) - E_{\min}) \quad , \quad (3.3)$$

Where both $E_{\max}$ and $E_{\min}$ remain unchanged with this transformation. These transformed energies should now display on average a constant level density. The spacing which have used in the determination of NNS distributions are given by

$$S_i = E'_{i+1} - E'_i \quad , \quad s_i = S_i / D \tag{3.4}$$

$D$ is the average of the spacing between the corrected energy levels. Now, with the spacing evaluated for each sequence, the NNS distribution is determined. For nuclear systems with time reversal symmetry which spectral spacing follows Gaussian Orthogonal Ensemble (GOE) statistics, the NNSD probability distribution function is well approximated by Wigner distribution[1-2]

$$P(s) = \frac{1}{2}\pi s e^{-\frac{\pi s^2}{4}} \quad , \tag{3.5}$$

While exhibits the chaotic properties of spectra. On the other hand, the NNSD of systems with regular dynamics is generically represented by Poisson distribution [1-2]

$$P(s) = e^{-s} \quad , \tag{3.6}$$

Different analyses which investigate the spectral statistics of nuclear systems propose a transitional behavior between these limits. To compare the fluctuation properties with regular and chaotic limits quantitatively, different distribution functions have been used [34-38]. One of popular distribution is Abul-Magd distribution [38] which was derived by assuming that, the energy level spectrum is a product of the superposition of independent subspectra, which are contributed respectively from localized eigenfunctions onto invariant (disjoint) phase space. The exact form of this model is complicated and its simpler form is proposed by Abul-Magd *et al* in Ref.[38] as:

$$P(s,q) = [1 - q + q(0.7 + 0.3q)\frac{\pi s}{2}] \times \exp(-(1-q)s - q(0.7+0.3q)\frac{\pi s^2}{4}) \quad , \tag{3.7}$$

Where interpolates between Poisson $(q = 0)$ and Wigner $(q = 1)$ distributions. In parametric estimation approaches, the value of distribution's parameter determine by different techniques which describe the chaotic or regular dynamics. To overcome the disadvantages of LSF-based estimated values, we have employed the Maximum Likelihood Estimation (MLE) technique [26] to estimate the parameter of distributions with more precision, i.e. estimated values yield accuracies which are closer to Cramer-Rao Lowe Bound (CRLB). The MLE procedure has described in detail in Ref.[26]. Here, we outline the basic ansatz and summarize the results.

### 3.1. The Maximum Likelihood-based results for Abul-Magd distribution

The MLE method provides an opportunity for estimating exact results with minimum variations. In order to estimate the parameter of distribution, Likelihood function is considered as product of all $P(s)$ functions [26],

$$L(q) = \prod_{i=1}^{n} P(s_i) = \prod_{i=1}^{n} [1 - q + q(0.7+0.3q)\frac{\pi s_i}{2}]\, e^{-(1-q)s_i - q(0.7+0.3q)\frac{\pi s_i^2}{4}} \quad , \tag{3.8}$$

Then, with taking the derivative of the log of likelihood function (3.8) respect to its parameter, $q$, and set it to zero, i.e., maximizing the likelihood function, the following relation for desired estimator is obtained

$$f:\sum\frac{-1+(0.7+0.6q)\frac{\pi s_i}{2}}{[1-q+q(0.7+0.3q)\frac{\pi s_i}{2}]}+\sum s_i+(0.7+0.6q)\frac{\pi s_i^2}{4} \quad , \tag{3.9}$$

We can estimate "$q$" by high accuracy via solving above equation by Newton-Raphson method which is terminated to the following result [26],

$$q_{new}=q_{old}-\frac{F(q_{old})}{F^{'}(q_{old})}=$$

$$=q_{old}-\frac{\sum\frac{-1+(0.7+0.6q_{old})\frac{\pi s_i}{2}}{[1-q_{old}+q_{old}(0.7+0.3q_{old})\frac{\pi s_i}{2}]}+\sum s_i-(0.7+0.6q_{old})\frac{\pi s_i^2}{4}}{\sum\frac{[0.3\pi s_i][1-q_{old}+q_{old}(0.7+0.3q_{old})\frac{\pi s_i}{2}]-[-1+(0.7+0.6q_{old})\frac{\pi s_i}{2}]^2}{[1-q_{old}+q_{old}(0.7+0.3q_{old})\frac{\pi s_i}{2}]^2}-\sum 0.15\pi s_i^2} \quad , \tag{3.10}$$

In ML-based technique, estimated parameters correspond to the converging values of iterations Eq.(3.10), where as an initial values we have chosen the values of parameters obtained by LSF method. On the other hand, the non-parametric estimation approach in dealing with NNSD, compare the histogram of each sequence with Poisson and Wigner curves [6-7]. This technique is unable to exhibit the intermediate statistics between limits. We used the Kernel Density Estimation (KDE) method [27] as an alternative to the histogram to describe the spectral statistics which interpolate between regular and chaotic dynamics via non-parametric estimation techniques, too. The aspects of KDE technique and its application in spectral investigation of nuclear systems are available in Ref.[27]. Here, we outline the basic ansatz and review the results.

### 3.2. Kernel Density Estimation (KDE)

In statistical application, Kernel Density Estimation (KDE) is regarded as a non-parametric technique for estimating the probability density function for sequence prepared by random variables. KDE is a fundamental data smoothing processes which inferences about the population are made based on a finite data sample. The simplest form of non-parametric D.E. is the familiar histogram. Assume $X_1, X_2, ...X_n$ are independent, identically distributed, real valued random variables with probability density $f$ . We consider estimators $\hat{f}$ of $f$ . We propose I($=I_k$) as partition of real line into disjoint intervals. If $h_k$ indicates the length of $I_k$, $N_k=\#\{i: X_i\in I_k, 1\le i\le n\}$ represents the number of observations in $I_k$ and $X=(X_1, X_2, ...X_n)$ we have [27]

$$\hat{f}(x)=\hat{f}_I(x,X)=\frac{N_k}{nh_k} \quad , \quad x\in I_k \tag{3.11}$$

$\hat{f}=\hat{f}_I$ is the histogram corresponding to selected partition and its length. The kernel estimator $\hat{f}=\hat{f}_a$ define as $(a=(K,h))$ [27]

$$\hat{f}(x) = \hat{f}_a(x, X) = \frac{1}{nh}\sum_{i=1}^{n} K(\frac{x - X_i}{h}) \quad , \quad (3.12)$$

Where $K$ is a kernel, namely a non-negative real function which integrates to one and $X_i$ is assumed as each member of sequence, i.e. $s_i$. We have used in this analysis the Gaussian kernel

$$K(\frac{x - X_i}{h}) = \frac{1}{\sqrt{2\pi}} e^{-\frac{(\frac{x-X_i}{h})^2}{2}} \quad , \quad (3.13)$$

Which explore the best efficiency in compare to other kernel functions. On the other hand, with using a simple bandwidth formula suggested by Scott [28], we determined the bandwidths for considered sequences which describe the smallest uncertainties (we represent these quantities in Figures (1-4)). To investigate chaotic or regular dynamics of nuclear spectra with KD-based estimated density function, we calculate the distance of $\hat{f}(x)$ related to GOE (or Poisson) limit via Kullback-Leibler Divergence (KLD) measure which define as [27]

$$D_{KL}(P\|Q) = \sum_i P(X_i) \ln \frac{P(X_i)}{Q(X_i)} \quad , \quad (3.14)$$

where $P(X_i)$ represents our estimated distribution function by MLE or KLD techniques, Poisson or GOE distributions are regard as $Q(X_i)$. The KLD measure is a non-symmetric measure to exhibit the average of the logarithmic difference between the probability distributions $P(X_i)$ and $Q(X_i)$. If $D_{KL}(P\|Q) \to 0$, a closer adaptation appear between two probability distribution functions, therefore, a closer distances to Poisson or GOE limits, explore regular or chaotic dynamics of sequences, respectively.

## 4. Numerical results

In this section, we explore the significant differences in the spectral statistics for even and odd mass nuclei which classified in different sequences. To this aim, we controlled all nuclei in the $34 \leq A < 206$ mass region and selected nuclei in which the spin-parity $J^{\pi}$ assignments of at least five consecutive levels are definite. With using all the available empirical data for $2^+$ and $4^+$ levels of even- and also for $1/2^+, 3/2^+, 5/2^+$ levels of odd-mass nuclei [23-25], levels are combined in several ways to search for effects due to the intensity of pairing, mass and also the types of pairs on the spectral statistics.

Since, the investigation of the majority of short sequences yields an overestimation about the degree of chaoticity measured by the "$q$" (Abul-Magd distribution's parameter) or "$KLD_{GOE}$" measures which evaluate the distances of KD-based estimated function to GOE limit. Therefore, we examine a comparison between the amounts of "$q$" or "$\langle KLD \rangle_{GOE}$" in the different sequences. This means, the smallest value (or smaller $\langle KLD \rangle_{GOE}$ measure) explain more regular dynamics and vice versa.

As have explained extensively in Refs.[21-22], the intensity of pairing interaction is weaker for odd mass nuclei in compare to even-even ones. To investigate the chaocity degrees for these nuclei, we prepared eight sequences of nuclei introduced in Table 1.

Table [1]. Even mass and odd-mass nuclei which have used to prepare sequences.

| Sequences | Nuclei |
|---|---|
| Even- Even nuclei | $^{34}$S, $^{38}$Ar, $^{42}$Ca, $^{44}$Ca, $^{46}$Ca, $^{48}$Ti, $^{50}$Ti, $^{50}$Cr, $^{52}$Cr, $^{54}$Cr, $^{58}$Fe, $^{64}$Ni, $^{66}$Zn, $^{68}$Zn, $^{68}$Ge, $^{70}$Ge, $^{70}$Zn, $^{72}$Ge, $^{74}$Ge, $^{74}$Se, $^{76}$Se, $^{78}$Se, $^{82}$Se, $^{82}$Kr, $^{84}$Kr, $^{86}$Kr, $^{88}$Sr, $^{90}$Zr, $^{92}$Zr, $^{92}$Mo, $^{94}$Zr, $^{96}$Mo, $^{98}$Mo, $^{102}$Ru, $^{104}$Ru, $^{106}$Ru, $^{102}$Pd, $^{106}$Pd, $^{108}$Pd, $^{110}$Pd, $^{110}$Cd, $^{112}$Cd, $^{114}$Cd, $^{116}$Cd, $^{112}$Sn, $^{118}$Sn, $^{120}$Sn, $^{122}$Xe, $^{118}$Te, $^{124}$Te, $^{126}$Te, $^{140}$Ce, $^{140}$Ba, $^{144}$Sm, $^{150}$Gd, $^{152}$Gd, $^{154}$Gd, $^{156}$Gd, $^{158}$Dy, $^{164}$Yb, $^{172}$Yb, $^{168}$Er, $^{182}$W, $^{180}$Pt, $^{182}$Pt, $^{188}$Os, $^{190}$Os, $^{192}$Os, $^{192}$Hg, $^{196}$Hg, $^{198}$Hg, $^{200}$Hg, $^{206}$Pb |
| Odd-mass nuclei | $^{43}$Ca, $^{47}$Ti, $^{49}$Ti, $^{53}$Cr, $^{57}$Fe, $^{61}$Ni, $^{67}$Zn, $^{73}$Ge, $^{81}$Kr, $^{77}$Se, $^{87}$Sr, $^{91}$Zr, $^{95}$Mo, $^{99}$Ru, $^{105}$Pd, $^{111}$Cd, $^{117}$Sn, $^{125}$Te, $^{131}$Xe, $^{137}$Ba, $^{141}$Ce, $^{143}$Nd, $^{145}$Pm, $^{149}$Sm, $^{151}$Eu, $^{155}$Gd, $^{155}$Gd, $^{159}$Tb, $^{161}$Dy, $^{163}$Ho, $^{167}$Er, $^{169}$Tm, $^{173}$Lu, $^{179}$Hf, $^{183}$W, $^{187}$Re, $^{189}$Os, $^{191}$Ir, $^{195}$Pt, $^{199}$Hg, $^{197}$Au |
| Even mass nuclei in $50 \leq A < 100$ mass region | $^{50}$Cr, $^{52}$Cr, $^{54}$Cr, $^{58}$Fe, $^{66}$Zn, $^{68}$Zn, $^{68}$Ge, $^{70}$Ge, $^{72}$Ge, $^{74}$Ge, $^{76}$Se, $^{78}$Se, $^{82}$Kr, $^{84}$Kr, $^{90}$Zr, $^{92}$Zr, $^{96}$Mo, $^{98}$Mo |
| Even mass nuclei in $150 \leq A \leq 200$ mass region | $^{150}$Gd, $^{152}$Gd, $^{154}$Gd, $^{164}$Yb, $^{182}$W, $^{180}$Pt, $^{182}$Pt, $^{188}$Os, $^{190}$Os, $^{192}$Os, $^{192}$Hg, $^{196}$Hg, $^{198}$Hg, $^{200}$Hg |
| Nuclei with proton-proton pairs | $^{50}$Ti, $^{70}$Zn, $^{74}$Se, $^{86}$Kr, $^{92}$Mo, $^{98}$Mo, $^{118}$Te, $^{140}$Ce, $^{168}$Er |
| Nuclei with neutron-neutron pairs | $^{34}$S, $^{42}$Ca, $^{64}$Ni, $^{76}$Se, $^{74}$Ge, $^{140}$Ba, $^{156}$Gd, $^{158}$Dy, $^{172}$Yb |
| Nuclei with holes in neutron levels | $^{46}$Ca, $^{70}$Ge, $^{82}$Se, $^{94}$Zr, $^{112}$Sn, $^{168}$Yb, $^{206}$Pb |
| Nuclei with holes in proton levels | $^{38}$Ar, $^{58}$Fe, $^{88}$Sr, $^{114}$Cd, $^{144}$Sm, $^{154}$Gd, $^{168}$Er |

These sequences unfolded and then, analyzed via MLE and KDE methods. Table 2 represents the $\langle KLD \rangle_{GOE}$ measures and also the ML-based estimated values for Abul-Magd distribution's parameter for sequences introduced in Table 1.The NNSDs of these sequences presented in Figures1-4.

Table [2]. ML-based estimated values for Abul-Magd distribution's parameter and also KLD measure for sequences introduced in Table 1.

| $\frac{\text{statistical criterion}}{\text{sequence}}$ | $\frac{\langle q \rangle}{\text{Abul}-\text{Magd parameter}}$ | $\langle KLD \rangle_{GOE}$ |
|---|---|---|
| Even mass nuclei | $0.1901 \pm 0.1522$ | $1.5208 \pm 0.1247$ |
| Odd-mass nuclei | $0.4337 \pm 0.1708$ | $0.9580 \pm 0.1351$ |
| Even mass nuclei in $50 \leq A < 100$ mass region | $0.6315 \pm 0.2945$ | $1.4003 \pm 0.1003$ |
| Even mass nuclei in $150 \leq A < 200$ mass region | $0.3927 \pm 0.1950$ | $1.6691 \pm 0.0953$ |
| Nuclei with proton-proton pairs | $0.5412 \pm 0.1946$ | $1.2576 \pm 0.0811$ |
| Nuclei with neutron-neutron pairs | $0.3208 \pm 0.1443$ | $1.6484 \pm 0.1140$ |
| Nuclei with holes in neutron levels | $0.4005 \pm 0.1933$ | $2.0611 \pm 0.0749$ |
| Nuclei with holes in proton levels | $0.5421 \pm 0.2339$ | $1.0375 \pm 0.1081$ |

The $\langle KLD \rangle_{GOE}$ measures which determine the distance of KD-based estimated functions to GOE (chaotic) limit, propose similar statistics that suggested by ML-based estimated values for considered systems. Also the obvious reductions in the uncertainties of KDE-based estimated function (the uncertainties have evaluated with Mean Absolute Error method [27-28]) have occurred, therefore, we can conclude, the KDE-based function yield the closer density function to real and exact distribution of every sequences.

These results, namely, more regular dynamics for even-mass nuclei in compare to odd-mass ones may be interpreted that the pairing force between the single particle and collective degrees of freedom is weaker in odd-mass nuclei than even-mass nuclei.

Also, a comparison of spectral statistics for even mass nuclei in different mass regions performed. First sequence constructed of nuclei in $50 \leq A < 100$ mass region, i.e. with $N(\text{ or } Z) \sim 20 \text{ to } 50$ and second one with prepared by nuclei located in the $150 \leq A \leq 200$ mass region, namely with $N(\text{ or } Z) \sim 50 \text{ to } 82$. Our considered criteria, "$q$" and "$KLD$" suggest more regularity for heavier nuclei (nuclei in *pf*- shell region) in compared to lighter ones (nuclei in the *sd*-shell region) which reveal theoretical predictions about chaotic dynamics of lighter nuclei [11,26] .

To look for the effect of pair type on spectral statistics, we have examined the level statistics of nuclei with different types of pairs out of closed shells. Theoretical analysis [39-40] proposed a deviation to more regular dynamics for nuclei with neutron-neutron pairs (with full proton energy levels) in compare to nuclei with proton-proton pairs (with full neutron energy levels in the shell model configuration which our results suggest similar statistics. This means, nuclei with proton-proton pairs which Coulomb force reduces the effect of pairing force in energy spectra, show more chaoticity. These results may be interpreted, pairing and Coulomb forces are competing with each others to dominate the regularity or chaoticity characteristics of nuclear spectra, respectively.

On the other hand, Figure 4 explores a comparison between spectral statistics of some nuclei with unclosed shells. For this analysis, we used the configuration of shell model and classified nuclei in two groups, nuclei whose protons occupy levels completely but the neutron levels have some empty states and the second group of nuclei by completely occupied neutron levels. Our results, namely, more regular dynamics for nuclei with unfilled neutron levels can be considered as the effect of competition between pairing and Coulomb forces while the stronger Coulomb force in nuclei with unfilled proton levels suggest a more chaotic dynamics [21-22].

As have been achieved in Ref.[26], the ML-based estimated values and corresponding distribution functions, exhibit less chaoticity in compare to LSF-based estimated distribution. From these tables and figures, KLD measures confirm the more regularity even more than predicted by ML-based estimated values and therefore consider regular dynamics for nuclear systems more than prediction of other estimation methods. Also, we see the apparent regularity of even- even mass nuclei. Since, the majority of these nuclei are deformed ones where the more regular dynamics for them is confirmed which is known as AbulMagd-Weidenmuller chaoticity effect [41-43].

## 5. Conclusion and Summary

In the present paper, we considered even and odd-mass nuclei in the $34 \le A \le 206$ mass region to describe the effect of intensity of pairing interaction, mass, different types of pairs on the spectral statistics. KDE and MLE methods have used to investigate the chaocity degrees of considered systems in the NNSD statistics framework. Our results propose more regular dynamics for even mass nuclei in compare to odd-mass ones and also a deviation to chaoticity in the energy levels of nuclei with unclosed proton levels. These results may suggest weaker coupling between the single particle and collective degrees of freedom in even-mass nuclei. We have shown that the pairing interaction is the main reason for the regular behavior in spectral statistics of low-lying states of semi-magic nuclei. Partial conservation of seniority is proposed as the underlying mechanism for the effect of pairing in low energy spectra. Also, one can conclude, a competition between pairing and coulomb interactions derive the spectral statistics of considered systems to regularity or chaocity, respectively. It is also worth mentioning that the studied mechanism is not exclusive to the pairing force. Other integrable Hamiltonians can have similar effects, i.e. $SU(3)$, as the effect seen here is related to the fact that some specific integrable part of Hamiltonian is dominant at low energy.

References

[1]. M. L. Mehta, Random Matrices (San Diego, Academic Press, 2nd ed 1991).
[2]. T. A. Brody, J. Flores, J. P. French, P. A. Mello, A. Pandey & S. S. M. Wong, Rev. Mod. Phys 53, 385 (1981).
[3]. O. Bohigas, M. J. Giannoni, C. Schmidt, Phys. Rev. Lett, 52, 1 (1984).
[4]. M. J. Giannoni, A. Voros, J. Zinn-Justin (Editors) , Chaos and Quantum Physics (Les Houches Session L II, North Holland, 1989)
[5]. S. Raman, T.A. Walkiewicz, S. Kahane, E.T. Jurney, J. Sa, Z. Gacsi, J.L. Weil, K. Allaart, G. Bonsignori and J.F. Shriner, Jr., Phys. Rev. C 43, 521 (1991).
[6]. F. Haake, Quantum Signatures of chaos (Springer- Verlag, Heidelberg, 2001)
[7]. R. U. Haq, A. Pandev, O. Bohigas, Phys. Rev. Lett 48, 1086 (1982).
[8]. M. V. Berry, Proc. R. Soc. London, Ser. A 400, 229 (1985).
[9]. G. E. Mitchell, E. G. Bilpuch, P. M. Endt, J. F. Shriner jr, Phys. Rev. Lett 61, 1173 (1988).
[10]. J. F. Shriner jr, E. G. Bilpuch, P. M. Endt, G. E. Mitchell, Z. Phys. A 335, 393 (1990).
[11]. J. F. Shriner jr, G. E. Mitchell, and T. von Egidy, Z. Phys. A 338, 309 (1991).
[12]. D. Mulhall, A. Volya and V. Zelevinsky, Phys. Rev. Lett 85, 4016 (2000).
[13]. J.M.G. Gómez, K. Kar, V.K.B. Kota, R.A. Molina, A. Relaño, J. Retamosa, Phys. Rep 499, 103 (2011).
[14]. T. Guhr, H. A. Weidenmuller, Ann. Phys. (N.Y) 199, 412 (1990).
[15]. Declan Mulhall, Phys.Rev.C 80, 034612 (2009) and 83, 054321 (2011).
[16]. R. A. Molina, J. M. G. Gomez and J. Retamosa, Phys. Rev. C 63, 014311 (2001).
[17]. E. Caurier, J. M. G. Gomez, V. R. Manfredi, L. Salasnich, Phys. Lett. B 365, 7 (1996).
[18]. J. M. G. Gomez, V. R. Manfredi, L. Salasnich, E. Caurier, Phys. Rev. C 58, 2108 (1998).
[19]. M. S. Bae, T. Otsuka, T.Mizusaki, N. Fukunishi, Phys. Rev. Lett 69, 2349 (1992).
[20]. Alexander Volya, Vladimir Zelevinsky, and B. Alex Brown, Phys. Rev. C 65, 054312 (2002).
[21]. R. A. Molina, Eur. Phys. J A 28, 125 (2006).
[22]. J. M. G. Gomez, R. Molina and J. Retamosa, Capture Gamma-Ray Spectroscopy and Related Topics (2003), pp. 396 (DOI: 10.1142/9789812795151_0051).
[23]. Nuclear data sheets up to 2013.
[24]. Live chart, Table of Nuclides, (http://www-nds.iaea.org/relnsd/vcharthtml/VChartHTML.html).
[25]. National Nuclear Data Center (Brookhaven National laboratory), chart of nuclides. (http://www.nndc.bnl.gov/chart/reColor.jsp?newColor=dm)
[26]. M. A. Jafarizadeh, N. Fouladi, H. Sabri and B. R. Maleki, Nucl. Phys. A 890-891, 29 (2012).
[27]. M. A. Jafarizadeh, N. Fouladi, H. Sabri and B. R. Maleki, Indian Journal of Physics (DOI: 10.1007/s12648-013-0311-7).
[28]. D. W. Scott, Multivariate Density Estimation: Theory, Practice and Visualization, John Wiley & Sons, (2009).
[29]. D. J. Dean, M. Hjorth-Jensen, Rev. Mod. Phys 75, 607 (2003).
[30]. A. Poves, A. P. Zuker, Phys. Rep 70, 235 (1981).
[31]. J. Dukelsky, S. Lerma H., L. M. Robledo, R. Rodriguez-Guzman, and S. M. A. Rombouts, Phys. Rev. C 84 061301 (2011).
[32]. Christian Pfleiderer, Rev. Mod. Phys 81,1551 (2009).
[33]. A. Bohr and B. R. Mottelson , Nuclear structure, Vol II, Nuclear Deformation , Benjamin , New York (1975).
[34]. T . A . Brody, Lett. Nuovo Cimento 7, 482 (1973).
[35]. M.V. Berry, M. Robnik, J. Phys. A: Math. Gen 17, 2413 (1984).
[36]. M.A.Jafarizadeh , N.Fouladi,H.Sabri and B.R.Maleki submitted to publish (nucl-th/1210.4751).
[37]. A.Y. Abul-Magd , H.L. Harney, M.H. Simbel , H.A. Weidenmüller, Phys. Lett. B 579, 278 (2004).
[38]. A.Y. Abul-Magd *et al*, Ann. Phys. (N.Y.) 321, 560 (2006).
[39]. Xizhen Wu, Zhuxia Li, Yingxun Zhang, Renfa Feng, Yizhong Zhuo, and Jianzhong Gu, AIP Conf. Proc. 597, 327 (2001).
[40].Belabbas, M.; Fellah, M.; Allal, N. H.; Benhamouda, N.; Ami, I.; Oudih, M. R. Int. J. Mod. Phys. E 19, 1973 (2010).
[41]. Paar .V and Vorkapic.D, Phys.Lett.B 205,7 (1988).
[41]. Paar .V and Vorkapic.D, Phys.Rev.C 41, 2397 (1990).
[42]. A. Y. Abul-Magd and M. H. Simbel, J.Phys.G: Nucl. Part. Phys 22, 104 (1996).

# Figure caption

Figure1. NNSDs for Even-Even mass and odd-mass nuclei based on KDE method. Solid, dashed and dotted line represent the KD-based density function, Poisson and GOE curves respectively.

Figure2. Similar to Figure 1, NNSDs for Even-Even mass nuclei in two mass regions, nuclei in $50 \leq A < 100$ and nuclei in $150 \leq A \leq 200$ mass regions, based on KDE technique.

Figure3. Similar to Figure 1, NNSDs for nuclei with proton-proton and neutron-neutron pairs based on KDE method.

Figure4. Similar to Figure 1, NNSDs for nuclei with unfilled neutron levels (holes in neutron levels) and nuclei with unfilled proton levels (holes in proton levels) based on KDE manner.

Figure1.

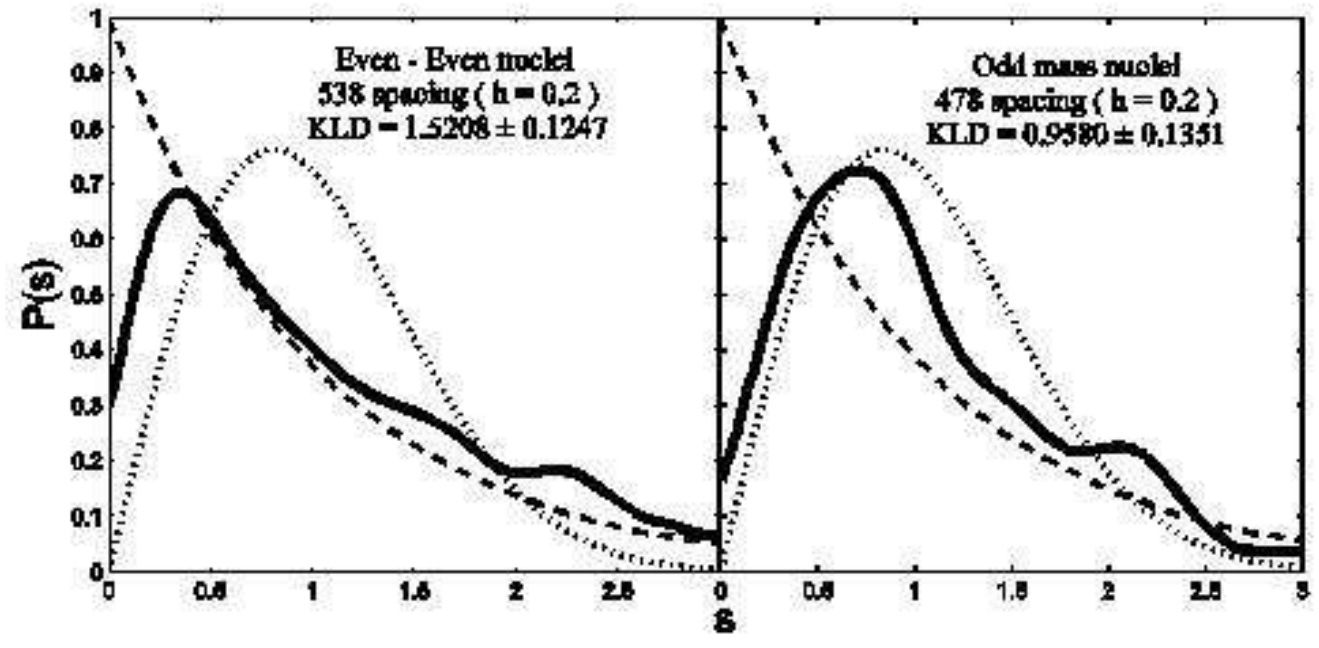

Figure2.

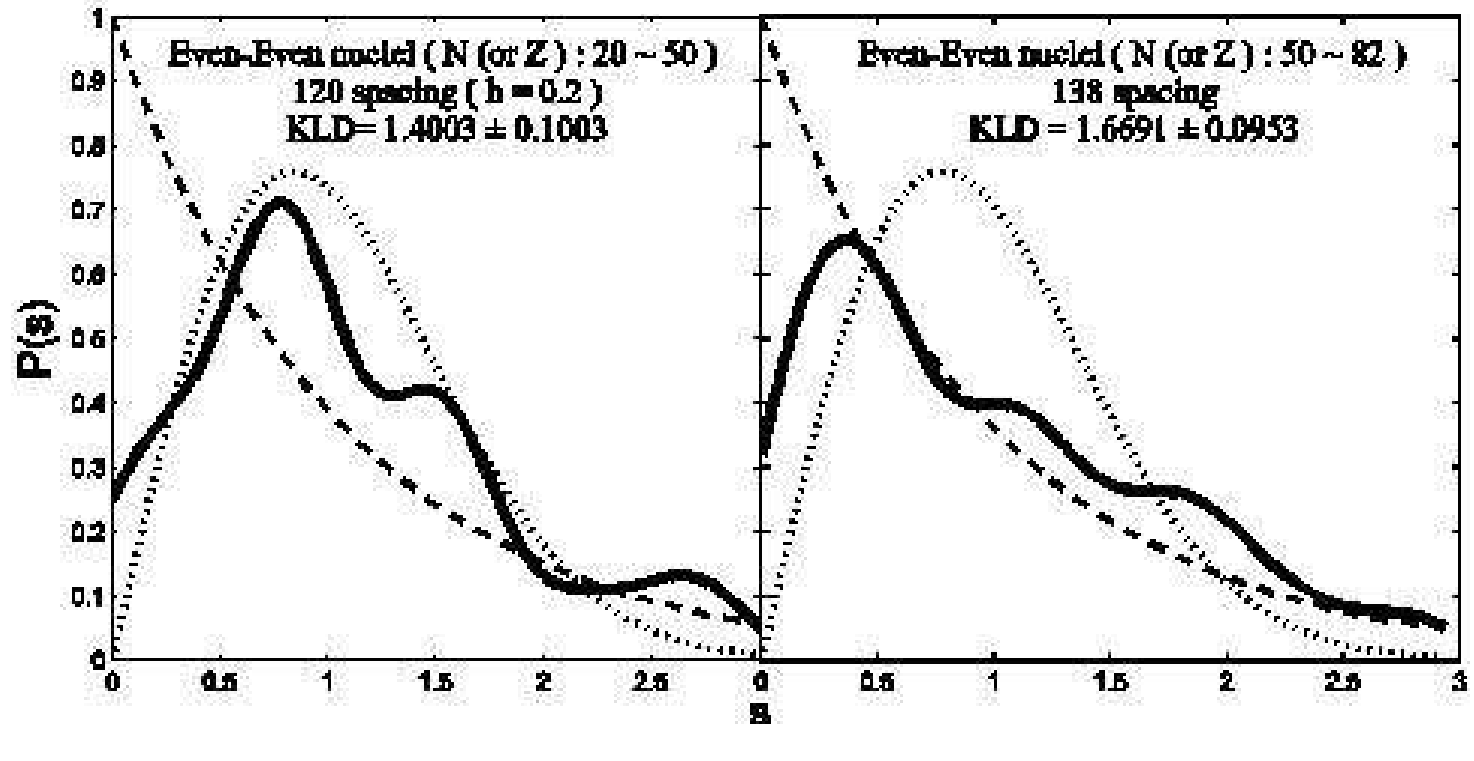

Figure3.

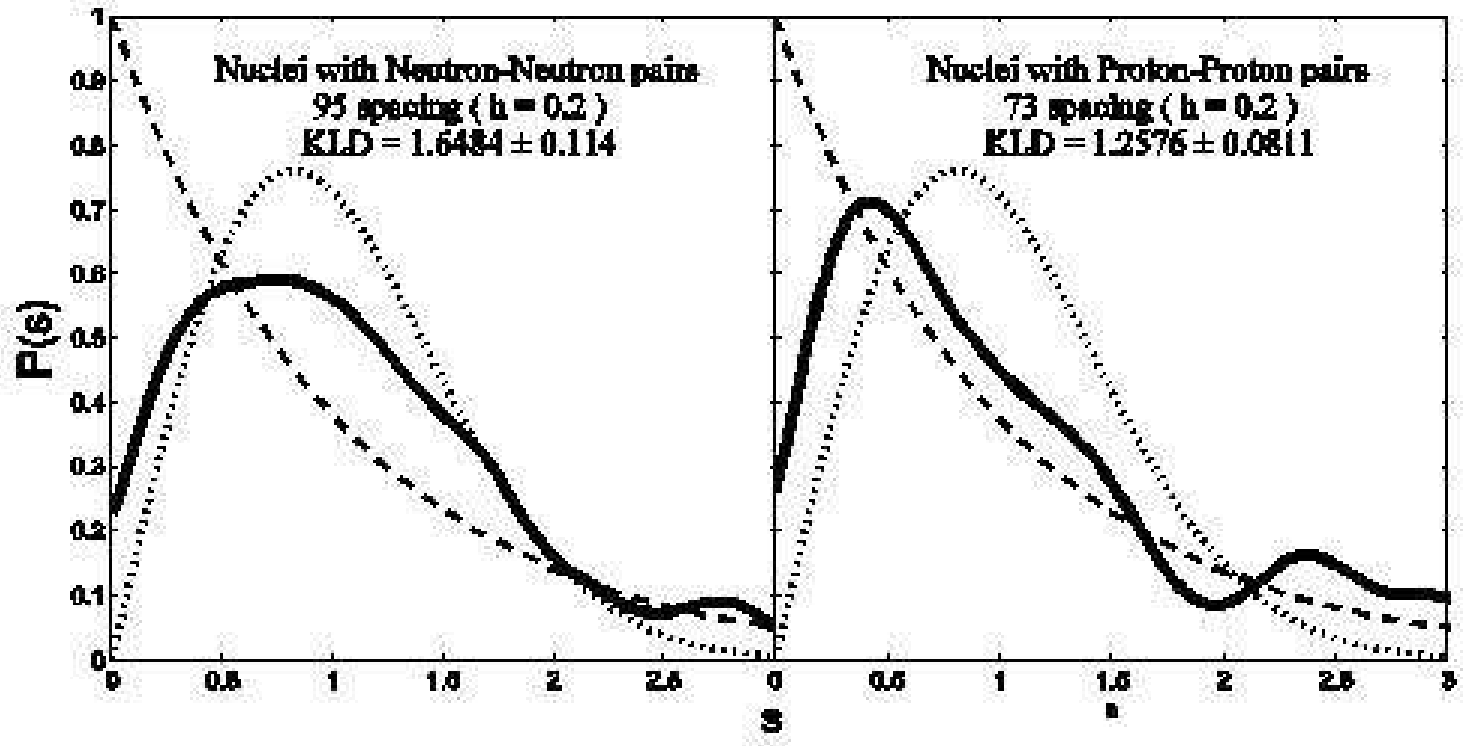

Figure4.

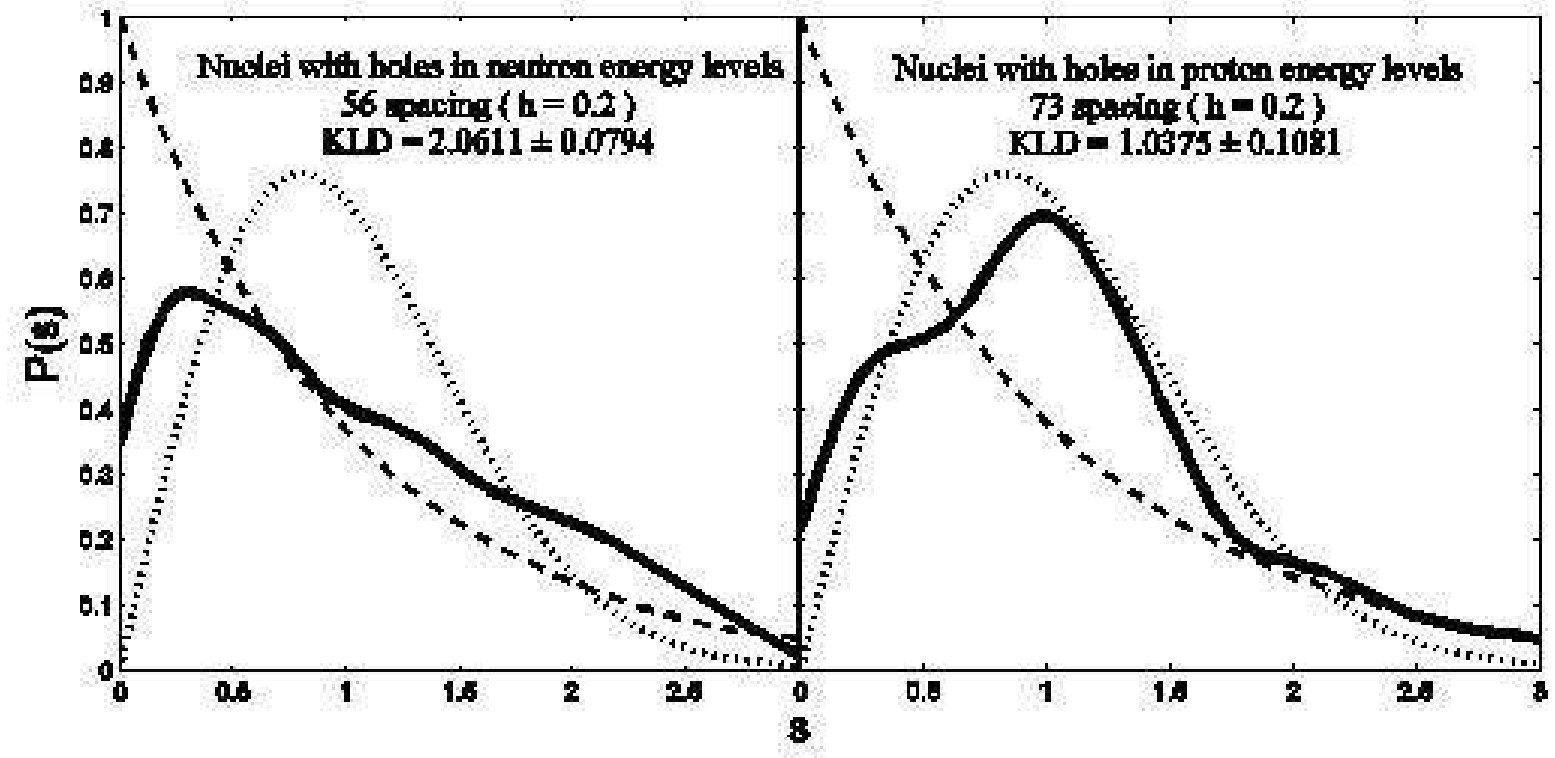